\documentclass{JHEP}
\usepackage{epsfig,graphics,lscape}
\usepackage{axodraw}
\renewcommand{\O}{{\mathcal{O}}}
\newcommand{\lag}{{\mathcal{L}}}
\newcommand{\leff}{\lag^{\mathit{eff}}}
\newcommand{\covder}{\!\not \! \!D}

\newcommand{\ltsim}{\lower3pt\hbox{$\, \buildrel < \over \sim \, $}}
\newcommand{\gtsim}{\lower3pt\hbox{$\, \buildrel > \over \sim \, $}}

\preprint{UG--FT--118/00 \\MIT-CTP-2997 \\hep-ph/0007316
\\ July 2000} 

\title{Observable contributions of new exotic quarks \\ to quark mixing}
\author{F. del Aguila$^a$,  M. P\'erez-Victoria$^b$ and J. Santiago$^a$ \\
 \vspace{0.5cm}
$^a$Departamento de F\'{\i}sica Te\'{o}rica y del Cosmos \\
Universidad de Granada \\
E-18071 Granada, Spain \\\vspace{0.5cm}
 $^b$Center for Theoretical Physics \\ Massachusetts
Institute of Technology \\ Cambridge, MA 02139, USA}
 
\date{\today}

\abstract{
Models with new vector-like quarks can produce observable quark mixing
effects which are forbidden in the Standard Model.
We classify all such models and write down the effective Lagrangian
that results from integrating out the new quarks. We study the
relations between neutral and charged currents and discuss how to
distinguish among the different possibilities. 
}

\begin{document}

\section{Introduction}

In a recent paper~\cite{letter1} we use an effective Lagrangian
approach~(see \cite{EffLag,Buchmuller,Arzt} and references there 
in) to
study the mixing of 
different flavours of quarks in trilinear couplings for general extensions 
of the Standard Model (SM). There we show
that, under quite general assumptions, only models containing
new exotic quarks near the TeV 
scale can produce large non-standard effects 
in the vertices $V\bar{q}q^\prime$ and $H\bar{q}q^\prime$, where $V$
is a massive gauge boson and $H$, the Higgs boson.
In this paper we study in detail the mixing of 
the known quarks in SM extensions that include any addition of
 vector-like
quarks. Our method will be to integrate out the new heavy fermions 
to obtain the
corresponding effective Lagrangian. The advantage of this approach is
that it isolates the interesting physics at the accessible energies
and allows us to discuss the different models with an arbitrary number
of exotic quarks in a unified
framework. Furthermore, we can use directly  the results in
Ref.~\cite{letter1}. Many aspects of quark mixing, mainly in extensions
with new vector-like singlet and doublet quarks, have been studied
in the literature~\cite{kane,papiro10}. Here we find new relations for
these 
models and consider isotriplets in detail
 in this context for the first time.

Let us first review the general effective description
of~\cite{letter1} and the arguments for the relevance of vector-like
quarks. Based on the success of precision tests of 
the SM, we assume that the only light fields are the ones in the
SM, with three families of quarks and leptons and
one Higgs doublet. Physics up to a scale 
$\Lambda$ is
described by the effective Lagrangian
\begin{equation}
\lag^{\mathit{eff}}
=\lag_4+\frac{1}{\Lambda^2}\lag_6+\ldots \, .
\label{leff}
\end{equation}
$\lag_4$ is the SM Lagrangian and $\lag_6$ describes corrections
arising from new physics at a scale $\Lambda$. The latter can be
written as 
\begin{equation}
\lag_6=\alpha_x{\mathcal{O}}_x+\; \mathrm{h.c.},
\end{equation}
where $\O_x$ is a gauge invariant operator of dimension 6 and we use
the repeated index convention unless otherwise specified.
The entire $\leff$ must be invariant under the 
$SU(3)_C\times SU(2)_L\times U(1)_Y$ gauge symmetry of the
SM. No $1/\Lambda$ terms are included in Eq.~(\ref{leff})
because when the almost exact lepton and baryon number
conservation is also imposed, dimension 5 operators involving the
SM fields are forbidden~\cite{Buchmuller}. The equations of motion of
$\lag_4$ allow to
reduce the number of operators $\O_x$ to 81, up to flavour indices. 

$\leff$ parametrises a large class of SM extensions, which includes
renormalizable gauge models with extra fermions, scalars and
gauge bosons, supersymmetric~\cite{susy} or not and in four 
or higher dimensions. The operators $\O_x$ are generated by the
exchange of the heavy degrees of freedom in the full theory. 
It is useful to distinguish those operators which are generated at
tree level from the ones which are only generated by loop
diagrams~\cite{Arzt}. If the high-energy theory is weak
interacting, which we assume henceforth, the coefficients
corresponding to loop-generated 
operators have an additional $1/16\pi^2$ suppression.
Therefore, large effects require that the new physics contributes at 
tree level. 
Here we are mainly interested in 
the vertices 
$V\bar{q}q^\prime$ and
$H\bar{q}q^\prime$. 
Processes with one of these bosons in the final state 
give direct information about these two-quark couplings.
The fact that four-fermion operators (of
dimension 6) do not contribute to these processes
greatly
reduces the number of operators to be considered: only 7 tree-level
generated operators are relevant~\cite{Arzt}. These
operators are listed in Ref.~\cite{letter1}.

After spontaneous symmetry breaking (SSB), 
the 7 operators give
corrections to $\lag_4$ of order $\alpha_x \frac{v^2}{\Lambda^2}$,
with $v\sim 250\, \mbox{GeV}$ the electroweak vacuum expectation
value (vev).   
If the high energy theory is a renormalizable gauge theory, the
operators that modify trilinear quark-gauge couplings can only be
generated at tree level by either   
extra gauge bosons mixing with the $Z$ or $W^\pm$ or extra quarks
mixing through Yukawa couplings with the SM ones.
However, the LEP data indicate that the mixing $\theta$ 
between SM and heavy gauge
bosons is typically  
 $\ltsim 0.01$~\cite{Langacker,LEP}, which gives an additional
suppression. Hence we are led to the conclusion that only extra quarks
can induce large new effects in quark mixing.

A fourth generation of chiral fermions is excluded at $99 \%$ C.L. by
the present limits on
 the S parameter~\cite{LEP}. On the other hand, vector-like fermions of
mass larger than the mass of the top are allowed.
In order to contribute to quark mixing, the 
new fermions must couple to the SM quarks (through
Yukawa couplings). This requires that they transform 
as  quarks under
$SU(3)_C$.
There are several possibilities for their
electroweak quantum numbers. In
Table~\ref{multiplets} we catalogue the exotic quarks $Q$ which can
mix with the SM fermions through a Yukawa coupling to the SM
scalar~\cite{papiro10}. Sometimes we shall refer to each of
these types of vector-like quarks as $Q^{(m)}$, where $m=1,\ldots,7$ 
indicates its
position in Table~\ref{multiplets}.
We stick to the SM
scalar sector because scalar singlets do not mix inequivalent
representations  and scalar triplets cannot have the
large vevs necessary to induce observable quark mixing, while if there
are 
additional scalar doublets we can consider the combination getting a
vev. 
Note that only one chirality of each
type of vector-like quark couples in this way to the SM fermions. In the
following we consider the addition of any number of these exotic
fermions to the SM. We integrate them out to obtain the corresponding
effective Lagrangian. Then we use the results in
Ref.~\cite{letter1} to write down the corrections to
the SM gauge and Yukawa couplings and find the relations and bounds
satisfied by the couplings in each model, which may allow to discriminate
between the different possibilities.

\begin{table}[!ht]
\caption{Vector-like quark multiplets $Q^{(m)}$ mixing with the SM quarks
through Yukawa 
couplings. The index $m$ labels the seven types of quark multiplet
additions in the given order.
The electric charge is the sum of the third component of
isospin $T_3$ and the hypercharge $Y$.} \label{multiplets}
\vspace{0.25cm}
\begin{tabular}{c|ccccccc}
$Q^{(m)}$ & $U$ & $D$ & $\left(\begin{array}{c} U\\D
\end{array}\right)$ &$\left(\begin{array}{c} X\\U
\end{array}\right)$ &$\left(\begin{array}{c} D\\Y
\end{array}\right)$ & $\left(\begin{array}{c} X\\U\\D
\end{array}\right)$ & $\left(\begin{array}{c} U\\D\\Y
\end{array}\right)$ \\ \hline
isospin & 0&0&$\frac{1}{2}$&$\frac{1}{2}$&$\frac{1}{2}$&1&1 \\
hypercharge & $\frac{2}{3}$&$-\frac{1}{3}$&$\frac{1}{6}$&
$\frac{7}{6}$& $-\frac{5}{6}$&$\frac{2}{3}$&$-\frac{1}{3}$
\end{tabular}\vspace{0.5cm}
\end{table}

\section{Effective Lagrangian for extensions with exotic quarks}
The complete Lagrangian for the quark sector is
\begin{equation}
\lag=\lag_l+\lag_h+\lag_{lh} \, , \label{lag}
\end{equation}
where $\lag_l$ and $\lag_h$ involve only light and 
heavy fields, respectively, and $\lag_{lh}$ contains their mixing. 
$\lag_l$ is the SM Lagrangian and coincides with $\lag_4$. Its quark sector
reads 
\begin{equation}
\lag_l^{\mbox{\tiny quark}} = \bar{q}^i_L i \covder q^i_L+\bar{u}^i_R
i \covder u^i_R+\bar{d}^i_R i \covder d^i_R - \left
( V^\dagger_{ij}\lambda^u_j\; \bar{q}^i_L \tilde{\phi}\; u^j_R 
+ \lambda^d_i\; \bar{q}^i_L \phi\; d^i_R \; +  \mbox{h.c.} \right) \, ,
\end{equation}
with $i,j$ running from 1 to 3. We assume without loss of generality
that the down Yukawa couplings $\lambda^d_i$ are diagonal. The up
Yukawas can be diagonalized after SSB by a rotation of $u_L$, producing the 
Cabibbo-Kobayashi-Maskawa (CKM) matrix $V$ in the charged currents. 
The scalar doublet has hypercharge $Y=1/2$ ($\phi$) or $Y=-1/2$ 
($\tilde{\phi}=\epsilon \phi^*$, with $\epsilon$ the antisymmetric
tensor of rank 2).
$\lag_h$ is quadratic in the heavy quarks:
\begin{equation}
\lag_h=\bar{Q}^a_L i\covder Q_L^a+\bar{Q}_R^a i\covder Q_R^a
- M_a (\bar{Q}_L^a Q_R^a+\bar{Q}_R^a Q_L^a) 
- (\tilde{\lambda}_{ab} \bar{Q}^a_L \Phi_{ab} Q^b_R 
 +\mbox{h.c.}),
\label{lag:h}
\end{equation}
with $a,b$ running from 1 to an arbitrary $n$. $Q^a$ can be any
multiplet in 
Table~\ref{multiplets}. 
For a given type of multiplet, $a$ and $b$ will also label the
different copies.
$\Phi_{ab}$ represents 
the appropriate form of the scalar required by gauge invariance. The
different possibilities are displayed in Table~\ref{Yukawasheavyheavy}.
The mass terms in Eq. (\ref{lag:h})
can be assumed to be
diagonal and real without loss of generality. These masses, which are
not protected by the gauge symmetry, determine the scale  in the 
effective Lagrangian: $\Lambda\sim M_a>v$. On the other hand, the
Yukawa couplings for the heavy quarks are non-diagonal and of order 1.
\begin{table}[!ht]
\caption{Values of $\Phi_{ab}$ for the different possibilities of
mixing between heavy fields.
\label{Yukawasheavyheavy}}
\vspace{0.25cm}
\begin{tabular}{c|cccc}
& $U^b_R $ &$D^b_R  $ & $\left(\begin{array}{c} X\\U\\D
\end{array}\right)^b_R $ & $\left(\begin{array}{c} U\\D\\Y
\end{array}\right)^b_R $ \\
\hline \\
$\overline{\left(\begin{array}{c}
U\\D \end{array}\right)_L^a}  $ &$\tilde{\phi}$ & $\phi$ &
$\frac{\sigma^I}{2} \tilde{\phi}$ & $\frac{\sigma^I}{2}\phi$  \\
$\overline{\left(\begin{array}{c} X\\U
\end{array}\right)_L^a}$ \rule{0cm}{1cm} 
& $\phi$ & $-$ & $\frac{\sigma^I}{2} \phi$ & $-$ \\
$\overline{\left(\begin{array}{c} D\\Y
\end{array}\right)_L^a}  $ \rule{0cm}{1cm} 
& $-$ & $\tilde{\phi}$ & $-$ & $\frac{\sigma^I}{2} \tilde{\phi}$
\end{tabular}\vspace{0.5cm}
\end{table}
Finally, $\lag_{lh}$
contains the Yukawa couplings mixing SM and new vector-like quarks. In
Table~\ref{mixingterms} we collect these terms for each type of
multiplet in Table~\ref{multiplets}. We use primes for these
Yukawa couplings and include $V$ in
the definition of some of them
 in order to simplify the final expressions. As we pointed out
only one chirality of each type of vector-like multiplet mixes. Because
$\lag_{lh}$ is linear in heavy fields, it can be written
\begin{equation}
\lag_{lh}=\bar{Q}^a_L\frac{\delta\lag_{lh}}{\delta\bar{Q}^a_L}+
\bar{Q}^a_R\frac{\delta\lag_{lh}}{\delta\bar{Q}^a_R}
+ \mbox{h.c.}, \label{lag:lh}
\end{equation} 
where $\frac{\delta\lag_{lh}}{\delta\bar{Q}^a }$ can be read from
Table~\ref{mixingterms}.
\begin{table}[!h]
\caption{Yukawa terms mixing light ($q^i_L,u^i_R,d^i_R$) and heavy
($Q^a_{L,R} $) quarks. The hermitian conjugate terms must be added. The
index I in the Pauli matrices corresponds to 
the $(+,0,-)$ basis of isospin, as
it does in the vector-like
triplets. 
The superscript $m=1,\ldots,7$ in $\lambda^{\prime(m)}$ labels
the different type of multiplet addition. 
$\frac{\delta\lag_{lh}}{\delta\bar{Q}^a_{L,R}}$ can be read 
directly from $\lag_{lh}$ which is linear in $\bar{Q}^a_{L,R}$.} 
\label{mixingterms}
\vspace{0.25cm}
\begin{tabular}{c|c}
$Q^{(m)}$ & $-\lag_{lh}$ \\\hline
$U$ \rule{0cm}{.6cm} & $\lambda^{\prime (1)}_{aj} V_{ji} 
\bar{U}^a_R\tilde{\phi}^\dagger q^i_L$ \\
$D$ \rule{0cm}{.6cm} & $\lambda^{\prime (2)}_{ai} \bar{D}_R^a \phi^\dagger 
\; q^i_L$ \\
$\left(\begin{array}{c} U\\D
\end{array}\right)$ \rule{0cm}{1cm}  & 
$
\lambda^{\prime (3u)}_{ai}
\overline{\left(\begin{array}{c} U\\D
\end{array}\right)_{L}^a} 
\tilde{\phi}  u^i_R 
+ \lambda^{\prime (3d)}_{ai}
\overline{\left(\begin{array}{c} U\\D
\end{array}\right)_{L}^a} 
\phi  d^i_R
$ \\
$\left(\begin{array}{c} X\\U
\end{array}\right)$ \rule{0cm}{1cm} &
$\lambda^{\prime (4)}_{ai}
\overline{\left(\begin{array}{c} X\\U
\end{array}\right)^a_{L}}
\phi  u^i_R$ \\
$
\left(\begin{array}{c} D\\Y
\end{array}\right)
$ \rule{0cm}{1cm}  & 
$\lambda^{\prime (5)}_{ai}
\overline{\left(\begin{array}{c} D\\Y
\end{array}\right)^a_{L}}
\tilde{\phi}  d^i_R
$ \\
$
\left(\begin{array}{c} X\\U\\D
\end{array}\right)
$ \rule{0cm}{1.2cm}  & 
$
\lambda^{\prime (6)}_{aj}
 V_{ji}
\overline{\left(\begin{array}{c} X\\U\\D
\end{array}\right)^a_{R}}_{\;I}
\tilde{\phi}^\dagger \frac{\sigma^I}{2} q^i_L
$\\
$
\left(\begin{array}{c} U\\D\\Y
\end{array}\right)
$ \rule{0cm}{1.2cm} &
$\lambda^{\prime (7)}_{aj}
 V_{ji}
\overline{\left(\begin{array}{c} U\\D\\Y
\end{array}\right)^a_{R}}_{\;I}
\phi^\dagger \frac{\sigma^I}{2} q^i_L
$
\end{tabular}\vspace{0.5cm}
\end{table}

\begin{figure}[h]
\begin{center}
\begin{picture}(300,80)(20,0)
\Line(45,40)(75,40)
\Line(45,42)(75,42)
\Line(25,30)(44,41)
\Line(95,30)(76,41)
\put(55,50){$\not \! p$}
\DashLine(45,40)(45,65){5}
\DashLine(75,40)(75,65){5}
\put(55,0){(a)}
\Line(155,40)(185,40)
\Line(155,42)(185,42)
\Line(135,30)(154,41)
\Line(205,30)(186,41)
\Photon(170,42)(170,80){3}{3}
\DashLine(155,40)(155,65){5}
\DashLine(185,40)(185,65){5}
\put(165,0){(b)}
\Line(265,40)(295,40)
\Line(265,42)(295,42)
\Line(245,30)(264,41)
\Line(315,30)(296,41)
\DashLine(280,42)(280,65){5}
\DashLine(265,40)(265,65){5}
\DashLine(295,40)(295,65){5}
\put(275,0){(c)}
\end{picture}\end{center}
\caption{Diagrams contributing to dimension 6 operators in the
effective Lagrangian. Light (heavy) quarks are depicted by a single
(double) line.} \label{diagrams}
\end{figure}
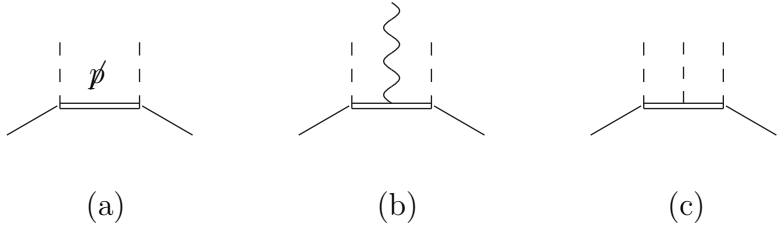

In order to find the
effective Lagrangian describing the physics below the scale $\Lambda$,
 we integrate out the heavy modes for a
generic addition of vector-like multiplets. 
Note that, unlike chiral fermions~\cite{feruglio}, vector-like quarks
decouple when their mass is sent to infinity. 
For our purposes it 
is sufficient to perform the integration at tree level, which can
be carried out imposing the equations of motion of the heavy modes. 
There is no need to
consider each kind of exotic quark separately, as $\lag_{lh}$ has
essentially the same form in all cases (see Ref.~\cite{letter1},
however, for a particular example). 
The requirement that the action be stationary under variations of the
heavy fields $\bar{Q}_{L,R}^a$
gives two coupled equations of motion (with no sum in $a$):
\begin{eqnarray}
i\covder Q^a_L - M_a Q^a_R - 
\tilde{\lambda}_{ab} \Phi_{ab} Q_R^b +
\frac{\delta\lag_{lh}}{\delta\bar{Q}^a_L}&=&0 \, , \\
i\covder Q^a_R - M_a Q^a_L - 
\tilde{\lambda}_{ab}^{\dagger} \Phi_{ab}^{\dagger} Q_L^b
+\frac{\delta\lag_{lh}}{\delta\bar{Q}^a_R}&=&0 \, ,
\end{eqnarray}
where we have used Eqs. (\ref{lag:h}) and (\ref{lag:lh}).
These equations can be solved iteratively. The
solution to order $1/M^2$ is
\begin{eqnarray}
Q^a_R&=&\frac{i\covder}{M_a^2}
\frac{\delta\lag_{lh}}{\delta\bar{Q}^a_R}  
+ \frac{1}{M_a} \frac{\delta\lag_{lh}}{\delta\bar{Q}^a_L} 
- \frac{1}{M_aM_b} \tilde{\lambda}_{ab} \Phi_{ab}
\frac{\delta\lag_{lh}}{\delta\bar{Q}^b_L} \, ,\\
Q^a_L&=& \frac{i\covder}{M_a^2}
\frac{\delta\lag_{lh}}{\delta\bar{Q}^a_L}  
+ \frac{1}{M_a} \frac{\delta\lag_{lh}}{\delta\bar{Q}^a_R} 
- \frac{1}{M_aM_b} \tilde{\lambda}_{ab}^{\dagger}\Phi_{ab}^{\dagger}
\frac{\delta\lag_{lh}}{\delta\bar{Q}^b_R} \, .
\end{eqnarray}
Inserting these expressions in the original Lagrangian in
Eq.(\ref{lag}) we obtain, to order $1/M^2$,
\begin{eqnarray}
\lag_4 & = & \lag_l, \\
\frac{1}{\Lambda^2} \lag_6 & = & 
\frac{\delta\lag_{lh}}{\delta Q^a_L} \frac{i\covder}{M_a^2}
\frac{\delta\lag_{lh}}{\delta\bar{Q}^a_L}
+
\frac{\delta\lag_{lh}}{\delta Q^a_R} \frac{i\covder}{M_a^2}
\frac{\delta\lag_{lh}}{\delta\bar{Q}^a_R}
- \left(\frac{\delta\lag_{lh}}{\delta Q^a_R} \frac{1}{M_aM_b} 
\tilde{\lambda}_{ab} \Phi_{ab} 
\frac{\delta\lag_{lh}}{\delta\bar{Q}^b_L} \; + \; \mathrm{h.c.}
\right)  \, . 
\label{Delta:lag:l}
\end{eqnarray}
We have set $\frac{\delta\lag_{lh}}{\delta Q^a_L}
\frac{\delta\lag_{lh}}{\delta \bar{Q}^a_R}=0$ since for each $a$ one of the
two factors vanishes.
Diagrammatically these dimension 6 corrections follow from the Feynman
diagrams in Fig.~\ref{diagrams}. The addition of the first two
diagrams generates the operators with a
covariant derivative, whereas the last diagram generates the operators
in parentheses in Eq.~(\ref{Delta:lag:l}).
%

In order to use the results of
Ref.~\cite{letter1} we have to write $\lag_6$ in terms of the operators of
Ref.~\cite{Buchmuller}.  The covariant derivative in
Eq.(\ref{Delta:lag:l}) acts on
$\frac{\delta\lag_{lh}}{\delta\bar{Q}^a}$, which is the product of 
the scalar and a light quark multiplet (see
Table~\ref{mixingterms}). Using the (covariant) Leibniz rule we
obtain terms in which the covariant derivative acts either on the
scalar field or on the light quark.
Whereas the former give---after a simple Fierz reordering of the
representation indices---operators in the catalogue of~\cite{Buchmuller},
the latter have to be transformed using the equations of motion of
$\lag_4$.  
The result takes the form (see Ref.~\cite{Buchmuller} for notation)
\begin{eqnarray}
\lag_6&=& 
(\alpha^{(1)}_{\phi q})_{ij} (\phi^\dagger iD_\mu \phi)(\bar{q}^i_L
\gamma^\mu q^j_L) 
+(\alpha^{(3)}_{\phi q})_{ij} (\phi^\dagger \sigma^I iD_\mu
\phi)(\bar{q}^i_L 
\gamma^\mu \sigma^I q^j_L) \nonumber \\
&&+(\alpha_{\phi u})_{ij} (\phi^\dagger iD_\mu \phi)(\bar{u}^i_R
\gamma^\mu u^j_R) 
+(\alpha_{\phi d})_{ij} (\phi^\dagger iD_\mu \phi)(\bar{d}^i_R
\gamma^\mu d^j_R) \nonumber \\
&&+(\alpha_{\phi \phi})_{ij} (\phi^T\epsilon iD_\mu \phi)(\bar{u}^i_R
\gamma^\mu d^j_R)
+
(\alpha_{u \phi })_{ij} (\phi^\dagger  \phi)(\bar{q}^i_L
\tilde{\phi} u^j_R) \nonumber \\
&&+(\alpha_{d \phi })_{ij} (\phi^\dagger  \phi)(\bar{q}^i_L
\phi d^j_R) + \mathrm{h.c.}\quad . \label{17}
\end{eqnarray}
Note that vector-like quarks generate only these 7 operators. 
In particular, operators of
magnetic-moment type or with stress-energy tensors do not appear at
this order. 
Four-fermion operators are not generated either, although one should keep
in mind that they may arise from other kinds of new physics.  
Each vector-like quark 
gives and independent contribution 
to one of
the first two terms in Eq.~(\ref{Delta:lag:l}), which is diagonal in
the heavy quark flavour, $a$. The term in parentheses in
this equation---which 
contributes only to $\O_{u\phi}$ and $\O_{d\phi}$---requires the
cooperative participation of two different types of vector-like
multiplets. 
Thus, the coefficients in Eq.~(\ref{17})
can be written
\begin{equation}
\alpha_x = \sum_m \alpha_x^m + \sum_{m<n} \alpha_x^{mn},
\end{equation}
where $\alpha_x^m$ is the diagonal contribution of the
 vector-like multiplets
$Q^{(m)}$ and $\alpha_x^{mn}$ is the  
combined contribution of multiplets $Q^{(m)}$ and $Q^{(n)}$. Note that
$\alpha_x^{mn}=0$ unless $x=u\phi,d\phi$, and only one of the two
quarks is a doublet.  
The contributions $\alpha_x^m$ and $\alpha_x^{mn}$ are collected in
Table~\ref{coefficients} and Table~\ref{mixedcoefficients},
respectively. 

\section{Corrections to quark couplings}

Upon SSB, the 7 operators in $\lag_6$ give 
$\frac{v^2}{\Lambda^2}$ corrections to
the SM Lagrangian. In the mass eigenstate basis these
corrections read:
\begin{eqnarray}
\lag^Z&=&-\frac{g}{2\cos \theta_W} \left( 
\bar{u}^i_L X^{uL}_{ij}
\gamma^\mu u^j_L+
\bar{u}^i_R X^{uR}_{ij}
\gamma^\mu u^j_R \right.\nonumber \\
&&\left.-\bar{d}^i_L X^{dL}_{ij}
\gamma^\mu d^j_L-
\bar{d}^i_R X^{dR}_{ij}
\gamma^\mu d^j_R - 2\sin^2\theta_W J^\mu_{EM}\right) Z_\mu, \nonumber \\
\lag^W&=&-\frac{g}{\sqrt{2}}(
\bar{u}^i_L W^{L}_{ij}
\gamma^\mu d^j_L+
\bar{u}^i_R W^{R}_{ij}
\gamma^\mu d^j_R)W^+_\mu +\textrm{h.c.}, \label{18} \\
\lag^H&=&-\frac{1}{\sqrt{2}}(
\bar{u}^i_L Y^{u}_{ij}
 u^j_R+
\bar{d}^i_L Y^{d}_{ij} d^j_R) H+ \mathrm{h.c.}, \nonumber 
\end{eqnarray}
with
\begin{eqnarray}
X^{uL}_{ij}&=& \delta_{ij}-\frac{v^2}{\Lambda^2}
V_{ik}(\alpha^{(1)}_{\phi q}-\alpha^{(3)}_{\phi
q})_{kl}V^\dagger_{lj}, \nonumber \\
X^{uR}_{ij}&=& -\frac{v^2}{\Lambda^2}
(\alpha_{\phi u})_{ij}, \nonumber \\
X^{dL}_{ij}&=& \delta_{ij}+\frac{v^2}{\Lambda^2}
(\alpha^{(1)}_{\phi q}+\alpha^{(3)}_{\phi
q})_{ij}, \nonumber \\
X^{dR}_{ij}&=& \frac{v^2}{\Lambda^2}
(\alpha_{\phi d})_{ij}, \label{19} \\
W^{L}_{ij}&=& \tilde{V}_{ik}(\delta_{kj}+\frac{v^2}{\Lambda^2}
(\alpha^{(3)}_{\phi
q})_{kj}), \nonumber \\
W^{R}_{ij}&=& -\frac{1}{2}\frac{v^2}{\Lambda^2}
(\alpha_{\phi \phi})_{ij}, \nonumber \\
Y^{u}_{ij}&=& \delta_{ij}\lambda^u_j-\frac{v^2}{\Lambda^2}
\left(V_{ik}(\alpha_{u \phi})_{kj} +
\frac{1}{4} \delta_{ij} [V_{ik}(\alpha_{u \phi})_{kj}+(\alpha_{u
\phi})^\dagger_{ik} V^\dagger_{kj}]\right)  ,\nonumber \\
Y^{d}_{ij}&=& \delta_{ij}\lambda^d_j-\frac{v^2}{\Lambda^2}
\left( (\alpha_{d \phi})_{ij}+\frac{1}{4}\delta_{ij} (\alpha_{d
\phi} +\alpha_{d \phi}^\dagger)_{ij} \right) , \nonumber  
\end{eqnarray}

\begin{table*}[!t]
\vspace{1cm}
\rotatebox{90}{\parbox{19cm}{
\caption{Coefficients $\alpha_x^m$ 
resulting from the
integration of an arbitrary number of each type of vector-like
quarks. The 
superscript $m=1,\ldots,7$ in $\lambda^{\prime(m)}$ labels
the different type of multiplet addition.}\label{coefficients}
}}
\vspace{1cm}
\hspace{.25cm}
\footnotesize
\leftskip 1.5cm
\rotatebox{90}{
\vspace{5cm}
\begin{tabular}{c|ccccccc}
$Q^{(m)}$ \rule[-.5cm]{0cm}{1cm} & $\frac{(\alpha^{(1)}_{\phi
q})_{ij}}{\Lambda^2}$  
& $\frac{(\alpha^{(3)}_{\phi q})_{ij}}{\Lambda^2}$ &
$\frac{(\alpha_{\phi u})_{ij}}{\Lambda^2}$ & 
$\frac{(\alpha_{\phi d})_{ij}}{\Lambda^2}$ 
&$\frac{(\alpha_{\phi \phi})_{ij}}{\Lambda^2}$ &
$\frac{(\alpha_{u \phi
})_{ij}}{\Lambda^2}$& $\frac{(\alpha_{d \phi})_{ij}}{\Lambda^2}$    
\\ \hline
$U$ \rule{0cm}{.8cm}
&$ \frac{1}{4}V^\dagger_{ik}\frac{\lambda^{\prime (1)\dagger}_{ka} 
\lambda^{\prime (1)}_{al}}{M^2_a} V_{lj}$ 
&$-\frac{(\alpha^{(1)}_{\phi q})_{ij}}{\Lambda^2}$  & $-$ & $-$ & $-$ &
$2\frac{(\alpha^{(1)}_{\phi q})_{ik}}{\Lambda^2} V^\dagger_{kj} 
\lambda^u_j$ &
$-$   
\\
$ D$ \rule{0cm}{.8cm}
&$ -\frac{1}{4}\frac{\lambda^{\prime (2) \dagger}_{ia} 
\lambda^{\prime (2)}_{aj}}{M^2_a}$ 
&$\frac{(\alpha^{(1)}_{\phi q})_{ij}}{\Lambda^2}$  & $-$ & $-$ & $-$ &
$-$& $-2\frac{(\alpha^{(1)}_{\phi q})_{ij}}{\Lambda^2} \lambda^d_j$ 
\\
$\left(\begin{array}{c} U\\D
\end{array}\right)$  \rule{0cm}{.8cm}
& $-$ & $-$ &
$ -\frac{1}{2}\frac{\lambda^{\prime (3u) \dagger}_{ia} 
\lambda^{\prime (3u)}_{aj}}{M^2_a}$ 
&$ \frac{1}{2}\frac{\lambda^{\prime (3d) \dagger}_{ia} 
\lambda^{\prime (3d)}_{aj}}{M^2_a}$ 
&$ -\frac{\lambda^{\prime (3u)\dagger}_{ia} 
\lambda^{\prime (3d)}_{aj}}{M^2_a}$ 
&$-V^\dagger_{ik}\lambda^u_k\frac{(\alpha_{\phi u})_{kj}}{\Lambda^2}$
&$\lambda^d_i\frac{(\alpha_{\phi d})_{ij}}{\Lambda^2}$
\\
 $\left(\begin{array}{c} X\\U
\end{array}\right)$  \rule{0cm}{.8cm}
& $-$ & $-$ &
$ \frac{1}{2}\frac{\lambda^{\prime (4)\dagger}_{ia} 
\lambda^{\prime (4)}_{aj}}{M^2_a}$ 
&$ -$
&$ -$
&$V^\dagger_{ik}\lambda^u_k\frac{(\alpha_{\phi u})_{kj}}{\Lambda^2}$
&$-$
\\
 $\left(\begin{array}{c} D\\Y
\end{array}\right)$  \rule{0cm}{.8cm}
& $-$ & $-$ &
$ -$
&$ -\frac{1}{2}\frac{\lambda^{\prime (5) \dagger}_{ia} 
\lambda^{\prime (5)}_{aj}}{M^2_a}$ 
&$ -$
&$-$
&$-\lambda^d_i\frac{(\alpha_{\phi d})_{ij}}{\Lambda^2}$
\\
 $\left(\begin{array}{c} X\\U\\D
\end{array}\right)$  \rule{0cm}{1cm}
&$ \frac{3}{16}V^\dagger_{ik}\frac{\lambda^{\prime (6) \dagger}_{ka} 
\lambda^{\prime (6)}_{al}}{M^2_a} V_{lj}$ 
&$\frac{1}{3}
\frac{(\alpha^{(1)}_{\phi q})_{ij}}{\Lambda^2}$  & $-$ & $-$ & $-$ &
$\frac{2}{3}
\frac{(\alpha^{(1)}_{\phi q})_{ik}}{\Lambda^2} V^\dagger_{kj} \lambda^u_j$ &
$\frac{4}{3}
\frac{(\alpha^{(1)}_{\phi q})_{ij}}{\Lambda^2} \lambda^d_j$    
\\
 $\left(\begin{array}{c} U\\D\\Y
\end{array}\right)$  \rule{0cm}{1cm}
&$ -\frac{3}{16}V^\dagger_{ik}\frac{\lambda^{\prime (7) \dagger}_{ka} 
\lambda^{\prime (7)}_{al}}{M^2_a} V_{lj}$ 
&$-\frac{1}{3}
\frac{(\alpha^{(1)}_{\phi q})_{ij}}{\Lambda^2}$  & $-$ & $-$ & $-$ &
$-\frac{4}{3}
\frac{(\alpha^{(1)}_{\phi q})_{ik}}{\Lambda^2} V^\dagger_{kj} 
\lambda^u_j$ &
$-\frac{2}{3}
\frac{(\alpha^{(1)}_{\phi q})_{ij}}{\Lambda^2} \lambda^d_j$    
\end{tabular}}\vspace{0.5cm}
\end{table*} 
\begin{table}[!ht]
\caption{Coefficients $\alpha_x^{mn}$ resulting from the integration
of an arbitrary number of vector-like quarks of each type due to the
mixing between vector-like multiplets.
The 
superscript $m=1,\ldots,7$ in $\lambda^{\prime(m)}$  labels
the different type of multiplet addition.
\label{mixedcoefficients}}
\vspace{0.25cm}
\footnotesize
\begin{tabular}{c|cc}
$Q^{(m)},\; Q^{(n)}$ \rule[-.4cm]{0cm}{.8cm}
 & $ \frac{(\alpha^{mn}_{u\phi})_{ij}}{\Lambda^2} $& 
     $ \frac{(\alpha^{mn}_{d\phi})_{ij}}{\Lambda^2}$ 
\\ \hline
$U, \;\left(\begin{array}{c} U\\D
 \end{array}\right)$ \rule{0cm}{.8cm}
 &
$V_{ik}^\dagger \frac{\lambda^{\prime(1) \dagger}_{ka}
\tilde{\lambda}_{ab} \lambda^{\prime(3u)}_{bj}}{M_a M_b}$&
$-$
\\
$U, \;\left(\begin{array}{c} X\\U \end{array}\right)$ \rule{0cm}{.8cm}
 & 
$V_{ik}^\dagger \frac{\lambda^{\prime(1) \dagger}_{ka}
\tilde{\lambda}_{ab} \lambda^{\prime(4) }_{bj}}{M_a M_b}$ &
$-$
\\
$D, \;\left(\begin{array}{c} U\\D \end{array}\right)$ \rule{0cm}{.8cm}
 &
$-$&
$ \frac{\lambda^{\prime(2) \dagger}_{ia}
\tilde{\lambda}_{ab} \lambda^{\prime(3d)}_{bj}}{M_a M_b}$
\\
$D, \;\left(\begin{array}{c} D\\Y \end{array}\right)$ \rule{0cm}{.8cm}
 &  $-$ &
$\frac{\lambda^{\prime(2)\dagger}_{ia}
\tilde{\lambda}_{ab} \lambda^{\prime(5)}_{bj}}{M_aM_b}$
\\
$\left(\begin{array}{c} U\\D
     \end{array}\right),\;\left(\begin{array}{c} X\\U\\D
     \end{array}\right) $ \rule{0cm}{1cm} &
$\frac{1}{4} 
V_{ik}^\dagger \frac{\lambda^{\prime(6) \dagger}_{ka}
\tilde{\lambda}_{ab} 
\lambda^{\prime(3u)}_{bj}}{M_aM_b}$
&
$ \frac{1}{2}
V_{ik}^\dagger \frac{\lambda^{\prime(6)
        \dagger}_{ka}\tilde{\lambda}_{ab}  
\lambda^{\prime(3d)}_{bj}}{M_aM_b}$
\\
$\left(\begin{array}{c} U\\D \end{array}\right)
,\;\left(\begin{array}{c} U\\D\\Y \end{array}\right)$ \rule{0cm}{1cm}
 & 
$\frac{1}{2}
V_{ik}^\dagger\frac{ \lambda^{\prime(7) \dagger}_{ka}
\tilde{\lambda}_{ab} 
\lambda^{\prime(3u)}_{bj}}{M_aM_b}$&
$ \frac{1}{4}
V_{ik}^\dagger\frac{ \lambda^{\prime(7) \dagger}_{ka}
\tilde{\lambda}_{ab} 
\lambda^{\prime(3d)}_{bj}}{M_aM_b}$
\\
$\left(\begin{array}{c} X\\U \end{array}\right), \;
\left(\begin{array}{c}  X\\U\\D \end{array}\right) $ \rule{0cm}{1cm} &
$-\frac{1}{4} V_{ik}^\dagger \frac{\lambda^{\prime(6) \dagger}_{ka}
\tilde{\lambda}_{ab} \lambda^{\prime(4)}_{bj}}{M_aM_b}$
& $-$
\\
$\left(\begin{array}{c} D\\Y \end{array}\right),\;
\left(\begin{array}{c} U\\D\\Y \end{array}\right)$ \rule{0cm}{1cm} &
$-$&
$-\frac{1}{4} V_{ik}^\dagger \frac{\lambda^{\prime(7) \dagger}_{ka}
\tilde{\lambda}_{ab} \lambda^{\prime(5)}_{bj}}{M_aM_b}$
\end{tabular}\vspace{0.5cm}
\end{table}
where we have introduced the unitary matrix
\begin{equation}
\tilde{V}=V+\frac{v^2}{\Lambda^2}(V A^d_L-A^u_L V).
\end{equation}
$A_L^{u,d}$ are the antihermitian matrices which, together with
$A_R^{u,d}$, diagonalize the corrected 
mass terms:
\begin{eqnarray}
(A^u_L)_{ij} & = & \frac{1}{2}(1-\frac{1}{2}\delta_{ij}) 
\frac{\lambda^u_i (V
\alpha_{u\phi})^\dagger_{ij} +(-1)^{\delta_{ij}}
 (V\alpha_{u\phi})_{ij} \lambda^u_j} 
{(\lambda^u_i)^2-(-1)^{\delta_{ij}}(\lambda^u_j)^2} \, ,\\
(A^d_L)_{ij} & = & \frac{1}{2}(1-\frac{1}{2}\delta_{ij}) \frac{\lambda^d_i
(\alpha_{d\phi})^\dagger_{ij} +(-1)^{\delta_{ij}}
 (\alpha_{d\phi})_{ij} \lambda^d_j} 
{(\lambda^d_i)^2-(-1)^{\delta_{ij}}(\lambda^d_j)^2} \, . 
\end{eqnarray}
$\lag_6$ in Eq. (\ref{17}) does not generate
any other trilinear coupling. A possible derivative coupling to the
Higgs is forbidden in these models because the corresponding 
coefficients $\alpha_x$ are always
hermitian (see Ref.~\cite{letter1}). With Eq. (\ref{19}) and
Tables
\ref{coefficients} and \ref{mixedcoefficients}
we can answer phenomenological questions on quark
mixing in processes with a vector boson or a Higgs.

\section{Relations and Bounds}

As can be readily observed from Eq. (\ref{19}), this effective
Lagrangian gives mixing effects which are forbidden in the SM. In
general: 
\begin{itemize}
\item There are flavour changing neutral currents (FCNC) in
the gauge interactions at tree level, 
 as the GIM mechanism~\cite{GIM} does not apply:
$X^{uL}_{ij},X^{dL}_{ij} \neq \delta_{ij}$. 
\item There are right-handed (RH) neutral currents not proportional to
$J^\mu_{EM}$: $X^{uR}_{ij},X^{dR}_{ij} \neq 0$. 
\item The left-handed (LH) charged currents are not described 
by a unitary CKM matrix: $W^L_{ik} W^{L\, \dagger}_{kj}, 
W^{L\, \dagger}_{ik} W^L_{kj} \neq
\delta_{ij}$. 
\item There are RH charged currents: $W^R_{ij}\neq 0$
\item There are FCNC mediated by the Higgs boson:
$Y^{u,d}_{ij}\neq \sqrt{2}\delta_{ij}\frac{m^{u,d}_j}{v}$. 
\end{itemize}

At this point it is important to remember that, whereas the couplings
 among the five lightest quarks are known with good precision, the
 couplings involving the top are only known with a precision of tens per
 cent. These couplings will be measured at the $1\%$ level at future
 colliders~\cite{Beneke}, 
which sets the limit on new observable effects in
 top mixing. From our arguments above on large mixing effects it is
 then clear that only vector-like quarks (with  $M/\lambda^\prime \ltsim 1$
 TeV) can induce observable new mixing                             
 effects in processes involving the top where a gauge boson or the Higgs is
 detected. Such effects would be described by Eq.~(\ref{19}) and
 Tables~\ref{coefficients} and \ref{mixedcoefficients}.
 Conversely, these processes can only prove this kind of new
 physics~\cite{letter1}. 
 
 On the other hand, the couplings of the five lightest quarks are more
 constrained by present data and cannot be too large ( $|X_{uc}|\ltsim
 10^{-3}$  and  $|X_{ds}|\ltsim 4\times 10^{-5}$ for
 instance~\cite{papiro10}).  
Observable corrections to the couplings of the five
 lightest quarks can also arise from other kinds of new physics
 although, if present at the TeV scale, extra quarks 
 would give the leading contribution for non-negligible 
$\lambda^\prime$
 ($\lambda^\prime \gtsim \lambda_b$).
For these small effects one must also take into account radiative
 corrections of similar size~\cite{barenboim}. 
 Typically, the most
 sensitive measurements of these couplings involve mixing of neutral
 mesons. One should be cautious in using our results in those processes, as
 four-quark operators can contribute as well. These operators are
 not generated by extra quarks but may be induced by other kinds of new
 physics, like extra gauge bosons. Even if four-quark operators are
 present, the results here may be employed to put bounds on the extra
 quark parameters assuming the absence of strong cancellations between
 cubic and quartic couplings.
We observe that
large new
 effects in top mixing and the required small new effects in the
 mixings of the other quarks can occur simultaneously in a natural
 way if the mechanism responsible for the hierarchy
 of the SM fermion masses produces a similar hierarchy in the Yukawa
 couplings mixing light and heavy quarks.
\begin{table}[!ht]
\rotatebox{90}{\parbox{20cm}{
\caption{$Z\bar{q}q^\prime$, $W\bar{q}q^\prime$ and $H\bar{q}q^\prime$
couplings to be measured 
experimentally (see Eq. (\ref{18})) and relations they fulfil to order
$\frac{1}{\Lambda^2}$ for each type
of vector-like quark additions $Q^{(m)}$.
In the SM $X^{uL}_{ij}= X^{dL}_{ij}=\delta_{ij},
X^{uR}_{ij}= X^{dR}_{ij}=0, W^L_{ij}=V_{ij}, W^R_{ij}=0,
Y^{u,d}_{ij}=\sqrt{2}\delta_{ij}\frac{m^{u,d}_j}{v}$.}\label{relations}
}}
\vspace{-1.5cm}
\hspace{.25cm}
\scriptsize
 \leftskip .5cm
\rotatebox{90}{
\begin{tabular}{c|cccccccc}
$Q^{(m)}$ \rule[-.4cm]{0cm}{.8cm} & 
$X^{uL}_{ij}$ & $X^{uR}_{ij}$&
$X^{dL}_{ij}$&  $X^{dR}_{ij}$& $ W^L_{ij}$& $W^R_{ij}$ &
$\frac{Y^{u}_{ij}}{\sqrt{2}}$
&$\frac{Y^{d}_{ij}}{\sqrt{2}}$
\\ \hline
$U$ \rule{0cm}{.6cm}
& $W^L_{ik}W^{L\dagger}_{kj} $
& 0 & $\delta_{ij}$ & 0 & $ V^L(1-\frac{1}{\Lambda^2})$ 
& 0 
&$ (-\frac{1}{2}\delta_{ij}+(1+\frac{1}{2}\delta_{ij})
X^{uL}_{ij})\frac{m^u_j}{v}$  
& $\delta_{ij}\frac{m^d_j}{v} $
\\
$ D$ \rule{0cm}{.6cm}
&$\delta_{ij}$&0 & $W^{L\dagger}_{ik}W^{L}_{kj} $
& 0 & $ V^L(1-\frac{1}{\Lambda^2})$  & 0 
& $\delta_{ij}\frac{m^u_j}{v} $
&$ (-\frac{1}{2}\delta_{ij}+(1+\frac{1}{2}\delta_{ij})
X^{dL}_{ij})\frac{m^d_j}{v}$  
\\
 $\left(\begin{array}{c} U\\D
\end{array}\right)$  \rule{0cm}{.7cm}
&$\delta_{ij}$& $\frac{1}{\Lambda^2}$  & $\delta_{ij} $
& $\frac{1}{\Lambda^2}$  & $V^L_{ij}$ &
$ \pm \frac{1}{\Lambda^2}$  
& $\frac{m^u_i}{v}(\delta_{ij}-(1+\frac{1}{2}\delta_{ij})X^{uR}_{ij}) $
& $\frac{m^d_i}{v}(\delta_{ij}-(1+\frac{1}{2}\delta_{ij})X^{dR}_{ij}) $
\\
 $\left(\begin{array}{c} X\\U
\end{array}\right)$  \rule{0cm}{.7cm}
&$\delta_{ij}$& $ -\frac{1}{\Lambda^2}$  & $\delta_{ij} $
& 0 & $V^L_{ij}$ & 0 
& $\frac{m^u_i}{v}(\delta_{ij}+(1+\frac{1}{2}\delta_{ij})X^{uR}_{ij}) $
& $\frac{m^d_i}{v}\delta_{ij} $
\\
 $\left(\begin{array}{c} D\\Y
\end{array}\right)$  \rule{0cm}{.7cm}
&$\delta_{ij}$& 0 & $\delta_{ij} $
& $-\frac{1}{\Lambda^2}$  & $V^L_{ij}$ & 0 
& $\frac{m^u_i}{v}\delta_{ij} $
& $\frac{m^d_i}{v}(\delta_{ij}+(1+\frac{1}{2}\delta_{ij})X^{dR}_{ij}) $
\\
 $\left(\begin{array}{c} X\\U\\D
\end{array}\right)$  \rule{0cm}{.8cm}
& $2\delta_{ij}-W^L_{ik}W^{L\dagger}_{kj} $
& 0 & $-\delta_{ij}+2W^{L\dagger}_{ik}W^L_{kj}$ & 0 & 
$ V^L(1+\frac{1}{\Lambda^2})$  & 0 
&$ (-\frac{1}{2}\delta_{ij}+(1+\frac{1}{2}\delta_{ij})
X^{uL}_{ij})\frac{m^u_j}{v}$  
&$ (\frac{5}{2}\delta_{ij}-(1+\frac{1}{2}\delta_{ij})
X^{dL}_{ij})\frac{m^d_j}{v}$  
\\
 $\left(\begin{array}{c} U\\D\\Y
\end{array}\right)$  \rule{0cm}{.8cm}
& $-\delta_{ij}+2W^L_{ik}W^{L\dagger}_{kj} $
& 0 & $2\delta_{ij}-W^{L\dagger}_{ik}W^L_{kj}$ & 0 &
$ V^L(1+\frac{1}{\Lambda^2})$  & 0 
&$ (\frac{5}{2}\delta_{ij}-(1+\frac{1}{2}\delta_{ij})
X^{uL}_{ij})\frac{m^u_j}{v}$  
&$ (-\frac{1}{2}\delta_{ij}+(1+\frac{1}{2}\delta_{ij})
X^{dL}_{ij})\frac{m^d_j}{v}$  
\end{tabular}
}
\vspace{2.5cm}
\end{table}

Once the couplings $X_{ij}$, $W_{ij}$~\cite{Espriu:2000yr} and 
$Y_{ij}$~\cite{ASB} 
have been measured (see Ref.~\cite{Beneke} for a review), one
may wonder what information they give about new physics above the
electroweak scale. 
We have already argued that the observation of non-standard effects,
especially in the 
case of top mixing,  may
point to the existence of new vector-like quarks.
In this case, one would also like to discriminate among
the different possible models. 
For instance, if the only exotic quarks are up
quark isosinglets, 
$U^a$, then $\alpha^{(1)}_{\phi q} =-\alpha^{(3)}_{\phi q}$ and
$X^{dL}_{ij}=\delta_{ij}$, $X^{uL}_{ij}=W^L_{ik}W^{L\dagger}_{kj}$. On
the other hand, if they are down quark isosinglets, 
$D^a$, $\alpha^{(1)}_{\phi q}
=\alpha^{(3)}_{\phi q}$  and
$X^{uL}_{ij}=\delta_{ij}$, $X^{dL}_{ij}=W^{L\dagger}_{ik}W^{L}_{kj}$. 
These general type of
relations between charged and neutral currents are discussed in
Ref.~\cite{letter1}. Here we only have to 
insert the values of the coefficients
$\alpha_x$ given in Table~\ref{coefficients}. In Table~\ref{relations},
we collect the relevant information when only one type of exotic 
quark gives a sizable contribution. 

Remember that these
expressions are valid to order $\frac{1}{\Lambda^2}$.
We use the following notation in Table~\ref{relations}: 
$ \frac{1}{\Lambda^2}$
indicates that the coupling is $\sim
\frac{v^2}{\Lambda^2}\lambda^{\prime 2}$ and that it is given by 
a positive semidefinite matrix. Similarly, $ -\frac{1}{\Lambda^2}$
stands for 
a negative semidefinite correction, $\pm \frac{1}{\Lambda^2}$
indicates an indefinite correction, and
$ 1+\frac{1}{\Lambda^2}$ and $ 1-\frac{1}{\Lambda^2}$ denote the
identity plus a positive and a negative semidefinite contribution,
respectively. 
$W^L$ is the product of a unitary matrix, $V^L=\tilde{V}$, times a
hermitian one~\cite{letter1}. 
The different exotic quark additions can be discriminated
by comparing the relations in Table~\ref{relations} with the
experimental couplings of the quarks to the $Z$, $W^\pm$ and
$H$. 
These relations are lost in general if more than one kind of  
vector-like quark is allowed. An exception in which the relations
between neutral and charged currents are preserved
 is the case when the extra
quarks belong to two types of multiplets, and these are a singlet and
a doublet or a triplet and a doublet. The reason is that singlets and
triplets only contribute to left-handed currents, while doublets only
contribute to right-handed ones.

On the other hand, as discussed in Ref.~\cite{letter1}, these
couplings also satisfy certain inequalities. 
The relevant information comes from the fact that the coefficient
matrices $\alpha_x$ in Table~\ref{coefficients} are either positive
or negative semidefinite. Indeed they are proportional to
$\frac{\lambda^{\prime\dagger}_{ia}\lambda^{\prime }_{aj}}{{M^2_a}}$.
In the case of the addition of just one type of heavy multiplet, 
the most stringent bounds for the Z couplings
are~\cite{letter1,AASM}
\begin{eqnarray}
|X_{ij}|^2&\leq &X_{ii} X_{jj}, \label{20}\\
|\delta_{ij}-X_{ij}|^2&\leq &(1-X_{ii}) (1-X_{jj})\, .\label{21} 
\end{eqnarray}
These inequalities
are also satisfied for combinations of extra multiplets that preserve the
semidefiniteness of the coupling matrices. For 
$X^{uL}$ ($X^{dL}$), Eqs.~(\ref{20},\ref{21}) hold for any combination
of extra 
multiplets not containing $Q^{(7)}$ ($Q^{(6)}$). In the case of $X^{uR}$
($X^{dR}$), they hold for combinations not containing $Q^{(3)}$ and
$Q^{(4)}$ ($Q^{(3)}$ and $Q^{(5)}$) at the same time.
Other inequalities are
related to the diagonal couplings: 
\begin{eqnarray}
X^{uL}_{ii}& \leq & 1 \hspace{1cm} \mbox{for combinations without
$Q^{(7)}$}, \\
X^{dL}_{ii}& \leq & 1 \hspace{1cm} \mbox{for combinations without
$Q^{(6)}$}, \\ 
X^{uL}_{ii}& \geq & 1 \hspace{1cm} \mbox{for $Q^{(7)}$ (and
doublets)}, \\ 
X^{dL}_{ii}& \geq & 1 \hspace{1cm} \mbox{for $Q^{(6)}$ (and
doublets)}. 
\end{eqnarray} 
The corresponding inequalities for $X^R$ can be read directly from
the signs in Table~6. 

On the other hand, the $W$ couplings fulfil the relations 
\begin{eqnarray}
W^L_{ik}W^{L\dagger}_{ki},W^{L\dagger}_{ik}W^{L}_{ki} & \leq & 1
\hspace{1cm} \mbox{for combinations without triplets},
\label{24} \\
W^L_{ik}W^{L\dagger}_{ki},W^{L\dagger}_{ik}W^{L}_{ki} & \geq & 1
\hspace{1cm} \mbox{for combinations without singlets}.
\label{26} 
\end{eqnarray}
Furthermore,
\begin{equation}
|W^R_{ij}|^2 \leq X^{uR}_{ii} X^{dR}_{jj}.
\label{25} 
\end{equation} 
Eq. (\ref{24}) ((\ref{26})) follows from the 
negative (positive) semidefiniteness of the coefficient matrix
$\alpha^{(3)}_{\phi q}$~\cite{letter1}, while  
Eq. (\ref{25}) is a consequence
of the explicit form of the SM correction in Eq. (\ref{19}) and
Table~\ref{coefficients}. The possibility
in Eq.~(\ref{26}) is not usually emphasized, because isotriplets
had not been studied in this context and detail before. 

These relations and inequalities may allow to discriminate between
different vector-like extensions if new mixing effects are observed.
In particular, some clear signatures are:  
\begin{itemize}
\item Non-zero $X^R$ requires new doublets.
\item Non-zero $W^R$ requires new doublets $Q^{(3)}$.
\item $X^L_{ii}>1$ or $(W^L{W^L}^\dagger)_{ii}>1$ requires new
triplets.
\item $(W^L{W^L}^\dagger)_{ii}<1$ requires new singlets.
\end{itemize}
On the other hand, as was shown in Ref.~\cite{letter1}, these
relations and inequalities can be used to put limits that are
independent of the details of the model~\cite{AASM}.   

Summarizing, we have derived the effective Lagrangian for quark
mixing for any SM extension with exotic quarks (Eqs. (\ref{18}),
(\ref{19}) and Tables~\ref{coefficients} and~\ref{mixedcoefficients}). 
Due to the particular form of these corrections, the couplings of two
quarks to the $Z$, $W^\pm$ and $H$
fulfil characteristic relations and inequalities which may allow to
discriminate among them and to obtain stringent constraints.
In the particular case of
the top quark the deviations from the SM predictions must be large to
be observable ($\sim 1\%$). We have pointed out that corrections of
this size can only arise from vector-like quarks mixing with the SM
ones. Therefore, the results in this paper are especially relevant for
the physics of top mixing.

\section*{Acknowledgements}
It is a pleasure to thank J.A. Aguilar-Saavedra, F. Cornet,
J.L. Cort\'{e}s, F. Feruglio, J. Prades and F. Zwirner 
for discussions. JS thanks the Dipartimento di Fisica ``Galileo
Galilei'' and INFN Sezione di Padova for their warm hospitality. 
This work has been  partially supported by  
CICYT and Junta de Andaluc\'{\i}a. 
JS and MPV also acknowledge financial
support from MECD.

\end{document}